\documentstyle[12pt]{article} \textwidth=18cm \textheight=23cm
\begin{document}

\centerline {\large\bf Rest mass or inertial mass?}

{\bf R. I. Khrapko}

{\it Moscow Aviation Institute \\
4 Volokolamskoe Shosse, 125871, Moscow, Russia.\\
khrapko\_ri@hotmail.com} \medskip

\centerline {ABSTRACT}

\noindent Rest mass takes the place of inertial mass in modern physics
textbooks. It seems to be wrong. But this phenomenon is hidden away
by the facts that rest mass adherents busily call rest mass {\it mass},
not {\it rest mass}, and the word {\it mass} is associated with a measure
of inertia. This  topic  has been  considered  by the  author  in  the
article "What  is  mass?" [1,  2,  3]. Additional  arguments  to  a
confirmation of such a thesis are presented here.  \medskip

\parshape=5
8cm 9cm 8cm 9cm 8cm 9cm 8cm 9cm 12cm 5cm
``Einstein's theory of the universe, based on the
principle that all motion is relative, and showing that mass varies
with its velocity, while space-time is a fourth dimension.''\\
{\it A. S. Hornby et al.} [4]
\medskip

\noindent The end of 20-th century was marked by a great mish-mash of
definitions of mass. \medskip

\noindent {\bf 1. Rest mass} \medskip

\noindent All was clear in the beginning of the century when the theory
of relativity was not yet created. Mass, $m$, denoted something like
amount of substance or quantity of matter. And at the same time mass was
the quantitative measure of inertia of a body.

     Inertia of a body determines momentum $\hbox{\bf P}$ of the body at
given velocity $\hbox{\bf v}$ of the body, i. e. it is a proportionality
factor in the formula
$$\hbox{\bf P} = m\hbox{\bf v}. \eqno(1)$$
The factor $m$ is referred to as inertial mass.

     But mass as a measure of inertia of a body can be defined also by
the formula
                            $$\hbox{\bf F} = m\hbox{\bf a}: \eqno (2)$$

     By this formula, the more is mass, the less is the acceleration of
a body at given force. Masses $m$ defined by the formulae (1) and (2)
are equal because the formula (2) is a consequence of the formula (1)
if mass does not depend on time and speed.  Thus,
\begin{quote}
``mass is the quantitative or numerical measure of body's inertia, that
is of its resistance to being accelerated" [5].
\end{quote}

     The same value of mass can be measured by weighing a body, that is
by measuring of the attraction to the Earth or to any other given body
(which mass is designated $M$). Thus, the same mass $m$ appears in the
Newton gravitational law
                                 $$F = \frac{\gamma Mm}{r^2}, \eqno(3)$$
but here $m$ is referred to as gravitational (passive) mass. This fact
expresses an equivalence of inertial and gravitational masses. Due to
this equivalence, the acceleration due to gravity does not depend on
the nature and the mass of a body:
                                  $$g = \frac{\gamma M}{r^2}. \eqno(4)$$

     Thus,
\begin{quote}
``mass is the quantity of matter in a body. Mass may also be considered
as the equivalent of inertia, or the resistance offered by a body to
change of motion (i. e. acceleration). Masses are compared by weighting
them." [6].
\end{quote} \medskip

\noindent {\bf 2. Inertial mass at high speeds}

     However, the special theory of relativity has shown that no body
can be accelerated up to the speed of light because the acceleration of
a body decreases to zero when the speed of the body approaches the
speed of light, however large the accelerating force is. This implies
that inertia of a body increases to infinity when the speed of the body
tends to the speed of light, though the ``amount of substance" of the
body obviously remains constant.

     More correctly, special relativity has shown that the momentum
$\hbox{\bf P}$ of a body at any speed is parallel to velocity
$\hbox{\bf v}$. Therefore the formula $\hbox{\bf P} = m\hbox{\bf v}$ is
valid at large speeds, if the coefficient $m$, that is inertial mass,
is accepted to be increased with speed in the fashion:
                       $$m = \frac{m_0}{\sqrt{1 - v^2/c^2}}, \eqno(5)$$
where c is the speed of light. That is, the expression
  $$\hbox{\bf P} = \frac{m_0\hbox{\bf v}}{\sqrt{1-v^2/c^2}} \eqno(6)$$
is valid for the momentum of a body.

     In these formulae $m_0$ is the value of mass which was spoken
about in the beginning. For a determination of the value, the body
should be slowed down and, after it, the formula (1) or (2) must be
applied at small speed. The value received by this method is called
{\it rest mass}. This mass, by definition, does not vary on
accelerating a body. Therefore, the formulae (1), (2), (3) must be
written as follow: $\hbox{\bf P} = m_0\hbox{\bf v}$, $\hbox{\bf F} =
m_0\hbox{\bf a}$, $F = \gamma Mm_0/r^2$. However, for small speeds, due
to formula (5), inertial mass is equal to rest mass, $m = m_0$, and
consequently the record (1), (2), (3 ) is correct in the ``before
special relativity section".

     To emphasize the fact that inertial mass $m$ depends on speed it
is named {\it relativistic mass}: it appears to have different values
from points of view of various observers if the observers have relative
velocities. Meanwhile, there is a preferred value of inertial mass
$m_0$. This value is observed by an observer which has no velocity
relative to the body. Such a property of inertial mass is similar to
the property of time: observers which are in motion relative to a clock
measure longer time intervals then the time interval measured by an
observer relative to whom the clock is at rest. This time interval is
called the proper time.  Thus,
\begin{quote}
``mass is the physical measure of the principal inertial property of a
body, i. e., its resistance to change of motion. At speeds small
compared with the speed of light, the mass of a body is independent of
its speed. At higher speeds, the mass of a body depends on its speed
relative to the observer according to the relation:
                               $$m = \frac{m_0}{\sqrt{1 - v^2/c^2}}, $$
where $m_0$ is the mass of the body by an observer at rest with respect
to the body, $v$ is the speed of the body relative to the observer who
finds its mass to be $m$." [7].
\end{quote}

     If you wish to check up the formula (6), you should measure
velocity {\bf v} and momentum {\bf P} of a body. The momentum of a body
is measured by the following operation. A moving body is braked by a
barrier, and during its braking the force $\hbox{\bf F}(t)$ acting on the
barrier is measured. The initial momentum of the body, by definition, is
equal to the integral
                    $$\hbox{\bf P} = \int\hbox{\bf F}(t)dt. \eqno(7)$$
It is postulated that this integral does not depend on details of
braking, that is on a form of function $\hbox{\bf F}(t)$.

     We should notice that the formulas (5) and (6) remain valid for
object which has no rest mass, $m_0 = 0$, for example, for photon or
neutrino (if one assumes that rest mass of neutrino is equal to zero).
Such objects have inertial mass and momentum, but they should move with
the speed of light. It is impossible to stop them: they disappear if
being stopped. Nevertheless, despite their speed is constant, their
inertial mass appear to be different for various observers. However, in
the case of such objects, no preferred value of inertial mass exists.
Or, it is possible to say, the preferred value of inertial mass is
equal to zero.

     We have detected the increase of inertia of a body at large speed
by a reduction of its acceleration at large speed. Thus we have
referred to formula (2). And it is allowable. However, just because of
the increase of inertial mass with the body velocity, the formula (2)
can change its form. The point is that at fixed acceleration, a force
directed in parallel with the velocity should supply not only the
increase of speed of available mass
                       $$m = \frac{m_0}{\sqrt{1 - v^2/c^2}}. \eqno(5)$$
It should also supply an increase of mass:
     $$\hbox{\bf F} = \frac{d}{dt}\hbox{\bf P} =
\frac{d}{dt}\left(\frac{m_0\hbox{\bf v}}{\sqrt{1 - v^2/c^2}}\right) =
\frac{m_0\hbox{\bf a}}{\sqrt{(1 - v^2/c^2)^3}}. \eqno(8)$$

     The coefficient
                                 $$\frac{m_0}{\sqrt{(1 - v^2/c^2)^3}}$$
is called ``longitudinal mass" [8].

     If the force is perpendicular to the velocity and so does not
change speed and inertial mass of a body, the formula $F = ma$ does not
change its form:
$$\hbox{\bf F} = \frac{m_0\hbox{\bf a}}{\sqrt{1 - v^2/c^2}}.\eqno(9)$$

     Using this circumstance, R.Feynman put forward a simple
operational definition of inertial mass $m$. ``We may measure mass, for
example, by swinging an object in a circle at a certain speed and
measuring how much force we need to keep it in the circle." [9].

     When the force has an arbitrary direction, the proportionality
factor in formula (2) must be considered as a certain operator (tensor)
which transforms vector {\bf a} to vector {\bf F}: $\hbox{\bf F} = \hat
m\hbox{\bf a}$. The operator $\hat m$ depends on speed and a direction
of the velocity of a body and, generally speaking, changes a direction
of a vector. It is easy to accept. You see, velocity {\bf v} is a
property of a body, but a force {\bf F} acting at the body is an
external agent with respect to the body. It is clear that a result of
the influence of the force, that is an acceleration of a body, can
depend on a correlation between directions of the vectors {\bf F} and
{\bf v}.\medskip

\noindent {\bf 3. Gravitational mass at higher speeds}

     At the same time the general theory of relativity has shown that
not only inertia of a body, but also its weight increases with speed by
the law (5). Indeed, the acceleration due to gravity of a body falling
downwards with speed $v$ is, roughly speaking,
     $$g = \frac{\gamma M(1-v^2/c^2)}{r^2}. $$
So, the weight of the body, according to (8), is
     $$ F = \frac{m_0 g}{\sqrt{(1 - v^2/c^2)^3}} =
\frac{\gamma Mm_0}{r^2\cdot\sqrt{1 - v^2/c^2}}. $$

So that inertial mass is equivalent to gravitational mass at any
speed $v$ of a body.

     The exact formula for the acceleration can be received within the
framework of the general theory of relativity as it is shown in Sec. 8:
     $$g = \frac{\gamma M(1-v^2/c^2)}{r\cdot\sqrt{r(r-r_g)}},
     \qquad r_g=2\gamma M/c^2. \eqno(10) $$
This formula is a relativistic generalization of the formula (4).
\medskip

\noindent {\bf 4. Energy}

Furthermore, special relativity has shown that an increment of inertial
mass, $m - m_0$, multiplied on square of the speed of light is equal to
kinetic energy of a body:
                                      $$(m - m_0)c^2 = E_k. \eqno(11)$$
     \begin{quote}
``A result of the theory that mass can be ascribed to kinetic energy  is
that the effective mass of the electron should vary with its velocities
according to the expression
                               $$m = \frac{m_0}{\sqrt{1 - v^2/c^2}}. $$
This has been confirmed experimentally." [6].
\end{quote}

     Therefore if we attach a rest energy $E_0 = m_0c^2$ to a body at
rest, the complete energy $E = E_0 + E_k$ of a body appears to be
proportional to inertial mass:
                                                $$E = mc^2. \eqno(12)$$

     This famous Einstein formula proclaims an equivalence between
inertial mass and energy. The two, up to now, different concepts are
incorporated in a single one.

     Thus,
\begin{quote}
  ``the formula $E = mc^2$ equates a quantity of mass $m$ to a quantity
of energy $E$. The relationship was developed from the relativity
theory (special), but has been experimentally confirmed" [7].
\end{quote}

     We should notice that the formula (12), as well as formulae (5)
and (6), are valid for an object which has no rest mass and rest
energy, $m_0= 0$.

     If you wish to check up the formula (11) and simultaneously to
make sure that special relativity is valid, you must measure the
inertial mass and the rest mass of a moving body as it was explained
above and, besides this, you must measure kinetic energy of the body:
              $$E_k =\int \hbox{\bf F}(l)d\hbox{\bf l}.$$
Here $\hbox{\bf F}(l)$ is the force acting on the barrier during the
body braking and $\hbox{\bf F}(l)d\hbox{\bf l}$ is a scalar product of
the force {\bf F} and an infinitesimal vector $d\hbox{\bf l}$ of
displacement of the barrier. (See [10]).

     The formula (11) connects inertial mass, rest mass and kinetic
energy. Using formula (6), it is easy to connect inertial mass, rest
mass and momentum:
                                    $$m_0^2 = m^2 - P^2/c^2.\eqno(13)$$

     For zero rest mass particles we receive:
                 $$mc = P,\quad{\rm or}\quad E = Pc.$$
\medskip

\noindent {\bf 5. System of bodies}

     If several bodies are considered to be a system of bodies, then,
as is known, their momenta and their inertial masses are summed up. For
two bodies this take the form:
                $$\hbox{\bf P} =\hbox{\bf P}_1 + \hbox{\bf P}_2, \qquad
                                              m = m_1 + m_2. \eqno(14)$$
In other words, momentum and inertial mass are additive.

     The case of the rest mass is entirely different. Equations (13),
(14) imply that rest mass of a pair of bodies with rest masses
$m_{01}$, $m_{02}$ is equal not to the sum $m_{01} + m_{02}$ but to a
complex expression dependent on the momenta $\hbox{\bf P}_1,$
$\hbox{\bf P}_2$:
$$m_0 = \sqrt{\left(\sqrt{m_{01}^2+
P_1^2/c^2}+\sqrt{m_{02}^2+P_2^2/c^2}\right)^2-
(\hbox{\bf P}_1+\hbox{\bf P}_2)^2/c^2}.\eqno(15)$$

     Thus, rest mass is, generally speaking, not additive. For example,
a pair of photons each having no rest mass does have a rest mass if the
photons move in different directions while the pair has no rest mass if
the photons move in the same direction.

     Nevertheless, the three quantities, {\bf P}, $m$, $m_0$, satisfy
the conservation law. That is, they remain constant with time for a
closed system.

     However, it seems to be unsuitable to consider rest mass of a
system of bodies because of the nonadditivity of rest mass. It is
meaningful to speak only about a sum of rest masses of separate bodies
of system. So, when one speaks that ``rest mass of final system
increases in an inelastic encounter" [11], the rest mass after the
encounter is compared with the sum of rest masses of bodies before the
encounter, but not with the system rest mass which is conserved thanks
to the nonadditivity. Just so, when one speaks about the mass defect at
nuclear reactions, for example, at synthesis of deuterium, $p + n = D +
\gamma$, the sum of the rest masses of proton and neutron is compared
with the sum of the rest masses of deuterium and ${\gamma}$-quantum, but
not with the system rest mass determined by the formula (15).\medskip

\noindent {\bf 6. A comparison  of  masses}

     And here a problem arises. Which of the two masses, the rest mass
or the inertial mass, must we name by a simple word {\it mass},
designate by the letter $m$ without indexes, and recognize as a "main"
mass? It is not a terminological problem. A serious psychologic
underlying reason is present here.

     To decide which of the masses is the main mass let us repeat  once
again properties of both masses.

     Rest mass is a constant quantity for a given body and denotes
``amount of substance of a body". It corresponds to a rudimentary Newton
belief that the masses stayed constant. But, rest mass is not
equivalent to the energy of a moving object, is not equivalent to
gravitational mass, rest mass is nonadditive and is not used as a
characteristic of a system of bodies or particles. This last
circumstance prevents the conservation law displaying. Particles moving
with the speed of light have no rest mass. The operational definition
of rest mass of a particle assumes its deceleration up to a small speed
without use of an information about current condition of the particle.

     Inertial mass is the relativistic mass. Its value depends on
observer's velocity. Inertial mass is equivalent to energy and to
gravitational mass, Inertial mass is additive, it satisfies the
conservation law. The operational definition of inertial mass is based
on the simple formula $\hbox{\bf P}=m\hbox{\bf v}.$

     From our point of view, inertial mass has to be called mass and to
be designate $m$, as it is done in the present article. \medskip

\noindent {\bf 7. Underlying psychologic reason}

     Unfortunately, plenty of physicists considers the rest mass as a
main mass, designates it by $m$, instead of $m_0$, and discriminates
the inertial mass. These physicists agree, for example, that the mass
of gas which is at rest increases with temperature since its energy
increases with temperature. But, probably, there is a psychologic
barrier that prevents them from explaining this increase by an increase
of masses of molecules owing to the increase of their thermal speed.

    These physics sacrifice the concept of mass as a measure of
inertia, sacrifice the additivity of mass and the equivalence of mass
and energy to a label attached to a particle with information about
``amount of substance" because the label corresponds the customary
Newton belief in invariable mass. And so they think that a radiation
which, according to Einstein [12], ``transfers inertia between emitting
and absorbing bodies" has no mass.

     Now inertial mass is excluded from textbooks and from popular
science literature [11, 13, 14], but this phenomenon is hidden by the
fact that rest mass adherents busily call rest mass {\it mass}, not
{\it rest mass}, and the word {\it mass} is associated with a measure
of inertia.

     The main psychologic difficulty is to identify mass and energy
(which varies), to accept these two essences as one. It is easy to
accept the formula $E_0 = m_0c^2$ for a body at rest. But it is more
difficult to accept a validity of the formula $E = mc^2$ for any speed.
The famous Einstein relation between mass and energy, that is a symbol
of 20-th century, seems ``ugly" to L. B. Okun' [15].

     Rest mass adherents are not, probably, capable to accept an idea
of relativistic mass the same as early opponents of special relativity
could not accept the relativity of time. The lifetime of an unstable
particle varies with velocity as its inertial mass:
     $$\tau=\tau_0/\sqrt{1-v^2/c^2}.$$

     It is appropriate to quote here from Max Planck:
\begin{quote}
     `An important scientific innovation rarely makes its way by
gradually winning over and converting its opponents: it rarely happens
that Saul becomes Paul. What does happen is that its opponents
gradually die out and that the growing generation is familiarized with
the idea from the beginning: another instance of the fact that the
future lies with youth. For this reason a suitable planning of school
teaching is one of the most important conditions of progress in
science.' [16]
\end{quote}

Unfortunately, the great idea of relativistic mass is carefully isolated
from youth. Now the article [1, 2, 3] is rejected by editors of the
following journals: ``Russian Physics Journal", ``Kvant" (Moscow),
``American Journal of Physics", ``Physics Education" (Bristol). ``Physics
Today". The present paper has been rejected by ``Russian Physics
Journal", ``Kvant" (Moscow), ``American Journal of Physics". \medskip

\noindent {\bf 8. Schwarzschild space}

     Here we will arrive at the formula (10) considering Schwarzschild
space-time [17] with the interval:
$$ds^2=\frac{r-r_g}{r}c^2dt^2-\frac{r}{r-r_g}dr^2-
r^2(d\theta^2+\sin^2\theta d\varphi^2)$$

     We get the equations of radial geodesic lines from the formulae
using the connection coefficients $\Gamma_{jk}^i$:
     $$\frac{d^2t}{ds^2}+
\frac{r_g}{r(r-r_g)}\cdot\frac{dr}{ds}\cdot\frac{dt}{ds}=0,\eqno(16)$$
     $$\frac{d^2r}{ds^2}-
\frac{r_g}{2r(r-r_g)}\left(\frac{dr}{ds}\right)^2+
\frac{(r-r_g)c^2r_g}{2r^3}\left(\frac{dt}{ds}\right)^2=0.\eqno(17)$$

     First integral of the equation (16) is:
     $$\frac{r-r_g}{r}\cdot\frac{dt}{ds}=\epsilon=Const.\eqno(18)$$

     We will record now an expression for the acceleration a taking
into account (18) and the fact that relationships between distance $l$
and time $t$, on the one hand, and coordinates $r$, $t$, on the other,
are given by the formulae
     $$dl=\sqrt{\frac{r}{r-r_g}}dr,\qquad
d\tau=\sqrt{\frac{r-r_g}{r}}dt:$$
     $$a=\frac{d}{d\tau}\frac{dl}{d\tau}=
\sqrt{\frac{r}{r-r_g}}\cdot\frac{d}{dt}
\left(\frac{r}{r-r_g}\cdot\frac{dr}{dt}\right)=
\frac{1}{\epsilon^2}\cdot\sqrt{\frac{r-r_g}{r}}\cdot\frac{d^2r}{ds^2}.
$$

     In this way, we have expressed the acceleration $a$ in terms of
$d^2r/ds^2.$ Now we can use the equation (17) and then, having reverted
to $l$ and $t$, we can arrive at
     $$a=-\frac{r_g(c^2-v^2)}{2r\sqrt{r(r-r_g)}},\qquad
v=\frac{dl}{d\tau}.\eqno(10)$$\medskip

     I thank G.S.Lapidus and V.P.Visgin. Their attention has helped me
to improve this paper text.
\medskip

\noindent {\bf References}

1.  R. I. Khrapko, "What is Mass?", Physics - Uspekhi, 43 (12), (2000).

2.  R. I. Khrapko, "What is Mass?", Uspekhi Phizicheskikh Nauk,
    170 (12), 1363 (2000) (in Russian).

3.  R. I. Khrapko, "What is mass?",\\
    http://www.mai.ru/projects/mai\_works/index.htm, No. 2 (in Russian).

4.  A. S. Hornby, E. V. Gatenby, H. Wakefield, The Advanced Learner's
    Dictionary of Current English, Vol.3, p.48 (ISBN 5-900306-45-3(3)).

5.  McGraw-Hill encyclopedia of science \& technology, V. 10 (McGraw-Hill
    Book Company, New York, 1987), p. 488.

6.  Chambers's Technical Dictionary, (W. \& R. Chambers, Ltd., London,
    1956), p. 529.

7.  Van Nostrand's scientific encyclopedia (Van Nostrand Reinhold, New
    York, 1989), p. 1796.

8.  M. Jammer, Concept of mass (Harvard University Press, Cambridge-
    Massachusetts, 1961).

9.  R. P. Feynman et al., The Feynman Lectures on Physics, V. 1
    (Addison-Wesley, Massachusetts, 1963), p. 9-1.

10.  R. I. Khrapko et al., Mechanics (MAI, Moscow, 1993) (in Russian).

11. E. F. Taylor and J. A. Wheeler, Spacetime Physics (Freeman, San
    Francisco, 1966).

12. A. Einstein, "Ist die Tragheit eines Korpers von seinem
    Energiegehalt abhangig?" Ann. d. Phys. 18, 639, (1905).

13. R. Resnick et al., Physics, V.1 (Wiley, New York, 1992).

14. M. Alonso, E. J. Finn, Physics (Addison-Wesley, New York, 1995).

15. L. B. Okun', "The concept of mass (mass, energy, relativity)",
    Physics - Uspekhi, 32(7), p. 637, (1989).

16. M. Planck, The Philosophy of Physics (George Allen \& Unwin
    Ltd, London, 1936), p. 90.

17. L. D. Landau, E. M. Lifshitz, The Classical Theory of Fields
    (Pergamon, New York, 1975).
\end{document}